\documentclass[
aps,prd,
10pt,
superscriptaddress,
amsfonts,amssymb,amsmath,
preprintnumbers,
nofootinbib,
a4paper,
eqsecnum,
]{revtex4}

\usepackage{bm,amssymb,amsmath,amsfonts,amsbsy}

\def\square{\mathchoice\sqr54\sqr54\sqr{2.1}3\sqr{1.5}3} 
\def\sqr#1#2{{\vcenter{\vbox{\hrule height.#2pt\hbox{\vrule
width.#2pt height#1pt \kern#1pt\vrule width.#2pt}\hrule height.#2pt}}}}
\def\square{\mathchoice\sqr54\sqr54\sqr{2.1}3\sqr{1.5}3}

\begin{document}

\preprint{YITP-10-106}

\title{Conformal transformations and
Nordstr\"om's scalar theory of gravity\footnote{Prepared for the
Proceedings of YKIS2010 on ``Cosmology\ --The Next Generation--"}
}

\author{Nathalie Deruelle}
\affiliation{
APC, UMR 7164 du CNRS, Universit\'e Paris 7, 
75205 Paris Cedex13, France}
\author{Misao Sasaki}
\affiliation{Yukawa Institute for Theoretical Physics, Kyoto University,
Kyoto 606-8502, Japan}
\affiliation{Korea Institute for Advanced Study,
Seoul 130-722, Republic of Korea}

\date{23 December, 2010}

\begin{abstract}
~\\
As we shall briefly recall, Nordstr\"om's theory of gravity is 
observationally ruled out. It is however an interesting  example
 of non-minimal coupling of matter to gravity  and of the role of conformal
 transformations. We show in particular that they could be useful to extend 
manifolds through curvature singularities.
\end{abstract}
\maketitle

\section{Introduction}

Scalar-tensor theories of gravity, where gravity is described by a $4$-dimensional metric $g_{\mu\nu}$ together with a scalar field $\Phi$, have been thoroughly studied (see e.g. the seminal paper \cite{DamourEsp}). Among them two stand apart: General Relativity which describes gravity by a metric alone, and Nordstr\"om's theory \cite{Nord} which describes it by a scalar field in flat spacetime (for reviews, see \cite{Ravndal}). General Relativity stands apart as it is observationally unchallenged. Nordstr\"om's theory is interesting as a toy model which shares with  General Relativity the unique property to embody the strong equivalence principle \cite{DamourEsp}. It has in particular been extensively used to test analytical \cite{Bruneton}
 or numerical \cite{ShapTeuk} methods developped to study the two-body problem.

In this paper we dwell on the two facets of a whole class of Nordstr\"om's theories,
 which can be formulated either as  field theories in Minkowski spacetime, as 
Nordstr\"om originally did,   or, for one of them, as a purely geometric theory,
 as Einstein and Fokker showed \cite{EinFok}.
 In today's jargon Nordstr\"om's formulation is an ``Einstein frame"  description since matter is non-minimally coupled to gravity and Einstein-Fokker's  is a  ``Jordan frame" description. Conformal transformations between Jordan and Einstein ``frames" (that is, metrics) are currently extensively used in e.g. ``modified-gravity" theories, such as those based on $f(R)$ actions often invoked to model  dark energy (see e.g. \cite{deFel} for a review).
 However there is still some debate on the equivalence of the two formulations 
(see \cite{Veiled} and references therein) and the example of Nordstr\"om's theories may help clarify this point.

\section{Nordstr\"om's theories of gravity}

\subsection*{1. The field equations}

In Nordstr\"om's theories, gravity is described by a massless scalar field $\Phi(x^\mu)$ in Minkowski spacetime. Its action is 
\begin{equation}
S_{\rm g}=-{c^3\over 8\pi G}
\int d^4x\sqrt{-\ell}\,\ell^{\mu\nu}\partial_\mu\Phi\partial_\nu\Phi\,,
\label{nordaction}
\end{equation}
where $c$ and $G$ are the speed of light and Newton's constant; 
the  coefficients of the Minkowski metric and its determinant are 
$\ell_{\mu\nu}$ and $\ell$ in the coordinate system $x^\mu$ (and  
reduce to $\ell_{\mu\nu}=\eta_{\mu\nu}=(-1,+1,+1,+1)$ in cartesian 
coordinates $X^\mu=(cT,X^i)$). The function $\Phi(x^\mu)$ is dimensionless.

Consider an ensemble of particles with (inertial) masses $m$ and 
4-velocities $u^\mu\equiv{dx^\mu/d\tau}$  where $c\tau$, such  
that $\ell_{\mu\nu}u^\mu u^\nu=-c^2$, gives the proper length along
 their worldline. Their action is chosen in analogy to that of an
 electric charge coupled to an electromagnetic potential:
\begin{eqnarray}
S_{\rm m}=-\sum mc^2\int (1+F(\Phi))d\tau\,,
\label{ptaction}
\end{eqnarray}
where $F$ is an {\sl a priori} arbitrary function of the gravitational 
potential $\Phi$ (hence the existence of a whole family of Nordstr\"om's theories).
 The fact that the coupling constant between a particle and gravity is $m$
 (and not some independent ``charge") embodies the weak equivalence 
principle: all particles will fall the same way in a gravity field. 

Extremising $S_{\rm g}+S_{\rm m}$ with respect to variations
 $\delta\Phi$ yields the equation of motion for $\Phi$~:
\begin{equation}
\square\Phi=-{4\pi G\over c^4}{dF/d\Phi\over1+F}T_{\rm m}
\label{Phieq}
\end{equation}
where $\square$ is the flat Dalembertian and where $T_{\rm m}$ is the
 trace of the matter stress-energy tensor:
\begin{equation}
T^{\mu\nu}_{\rm m}\equiv-{2c\over\sqrt{-\ell}}
{\delta S_{\rm m}\over\delta \ell^{\mu\nu}}
=\sum mc\int (1+F){u^\mu u^\nu\over\sqrt{-\ell}}
\delta_4[x^\lambda-x^\lambda(\tau)])d\tau\,.
\label{Tmatter}
\end{equation}

As for the equations of motion of the particles they are obtained by 
extremizing the matter action with respect to variations $\delta x^\mu$ of their path:
\begin{equation}
{Du^\mu\over d\tau}
=-{c^2\over1+F}\left(\partial^\mu F+{u^\mu u^\nu\over c^2}\partial_\nu F\right)
\qquad\Longleftrightarrow
\qquad D_\nu T^{\mu\nu}_{\rm m}
={T_{\rm m}\over1+F}\,\partial^\mu F\,,
\label{ptmotion}
\end{equation}
where $D$ is the covariant derivative associated with $\ell_{\mu\nu}$.
 By construction they do not depend on $m$. 

\smallskip
The equations of motion when matter is an ensemble of particles suggest to generalize them,
whatever matter may be, into
\begin{equation}
\left\{\begin{aligned}
\square\Phi
&=-{4\pi G\over c^4}{dF/d\Phi\over1+F}T_{\rm m}\,,
\cr
D_\nu& T^{\mu\nu}_{\rm m}={T_{\rm m}\over1+F}\,\partial^\mu F\,. 
\end{aligned}\right.\label{eom}
\end{equation}
If matter is a scalar field $\Psi$ with potential $V(\Psi)$, 
the action and stress-energy-tensors are
\begin{equation}
\left\{\begin{aligned}
S_{\rm m}&=-\int{d^4x\over c}(1+F)^4
\sqrt{-\ell}\left[{\ell^{\mu\nu}\over (1+F)^2}
{1\over2}\partial_\mu\Psi\partial_\nu\Psi+V(\Psi)\right]\,,
\cr
T_{\mu\nu}^{\rm m}&=(1+F)^2\left(\partial_\mu\Psi\partial_\nu\Psi
-{1\over2}\ell_{\mu\nu}\partial_\rho\Psi\partial^\rho\Psi
-(1+F)^2\ell_{\mu\nu}V(\Psi)\right)\,.
\end{aligned}
\right.\label{Sm}
\end{equation}
If matter is a perfect fluid, 
\begin{equation}
T_{\mu\nu}^{\rm m}
=(1+F)^4\left[(\bar\epsilon+\bar p){u_\mu u_\nu\over c^2}
+\bar p\,\ell_{\mu\nu}\right]\,,
\label{Tperf}
\end{equation}
where $\bar\epsilon$ and $\bar p$ being its energy density and pressure.
See Appendix for a justification of the various couplings to $\Phi$.

\subsection*{2. Time and inertial mass in Nordstr\"om's gravity}

 In Special Relativity, {\sl in the absence of gravity}, Minkowskian 
coordinates $X^\mu$, where $\ell_{\mu\nu}=\eta_{\mu\nu}$, are postulated to 
 represent time and  an inertial frame, $T\equiv X^0/c$  being time as measured
 by a clock at rest in that frame. As for ``proper time" $\tau$,
 it is postulated to represent time as measured along the time-like 
worldline of a particle of inertial mass $m$.
  
However these postulates have to be revisited  in the presence of gravity.
Indeed, the field $\Phi$, being long range, pervades the whole universe and 
no particle escapes its grip since its ``charge" is its inertial mass.
 Therefore, outside asymptotic infinity where $\Phi$ may vanish, there are no
 longer free particles which can materialize an inertial frame and the 
statement that $T$ is the time measured by a clock at rest becomes inoperational. 

To decide which quantity must represent time, let us consider, for example, 
an electron, coupled to the electromagnetic potential $A_\mu$ of a proton say,
 and in the presence of a gravity field, which can be taken to be 
approximately constant around $x^\mu$. Its action  is
\begin{equation}
S_{\rm m}=- mc^2\int (1+F)d\tau+q\int A_\mu dx^\mu\,,
\end{equation}
where $m$ and $q$ are its inertial mass and charge, where 
$A_\mu=(-q/r,\vec0)$ if the proton is considered at rest, 
and where $d\tau\equiv\sqrt{1-{V^2/c^2}}\,dT$  with
 $V^\alpha\equiv{dX^\alpha/dT}$ in Minkowskian coordinates
 $X^\mu=(cT,X^\alpha$).
Now, the infinitesimal element $m(1+F)d\tau$ can be rewritten in 
two different equivalent ways as
\begin{eqnarray}
m(1+F)d\tau
&=&m\,d\bar\tau\qquad\mbox{with}\qquad d\bar\tau\equiv (1+F)\,d\tau\,,
\label{wtime}
\\
\mbox{or}\hspace*{5mm}
m(1+F)d\tau&=&\bar m\,d \tau\qquad\hbox{with}\qquad 
\bar m\equiv (1+F)\,m\,.
\label{wmass}
\end{eqnarray}
Therefore, as most clearly stated by Dicke \cite{Dicke},
 all predictions of Maxwell's classical theory will hold in the presence of gravity,
 if either
\\
\noindent
(a) time is postulated to be represented  by $\bar\tau$, and no longer
  by  $\tau$:  thus time measured by a clock at rest in a frame where 
the gravity field is constant and where electrically neutral particles
 have a uniform motion, is not $T$ but the rescaled time, 
or ``Weyl time", $\bar T=(1+F) T$, 
\\
\noindent
(b) or elementary particles are endowed with a varying effective mass
 $\bar m$, $\tau$ remaining their proper time along their wordline, 
see e.g. \cite{Veiled} and references therein.

We examine now a few consequences of these two new, alternative, postulates.

\subsection*{3. Nordst\"om's theories and Solar System observations}

\noindent
$\bullet$ {\sl Bending of light.}

As in Newton's theory the status of particles with zero 
mass is {\sl a priori} ambiguous~: either they can be
 supposed  to obey (\ref{ptmotion}) or to travel at $c$
 along Minkowskian light cones.
However it is easy to see, by solving (\ref{ptmotion})
 at lowest order in the gravitational field of a central 
object ($F(\Phi)=\Phi+{\cal O}(\Phi^2)$ with 
$\Phi=-{\cal M}/r$ where ${\cal M}\equiv GM/c^2$), 
that ultra-relativistic particles are not deflected. 
Hence  the Nordstr\"om 
value of the PPN parameter $\gamma$, defined 
by $(1+\gamma)/2={\Delta\phi/\Delta\phi_{\rm GR}}$, 
is $\gamma=-1$. This prediction does not involve time measurements
 and hence does not involve the new postulates.

Since deviation of light has been measured and found to 
agree with the GR prediction ($\gamma=1$), this is enough to rule
 out all of Nordstr\"om's theories, whatever $F(\Phi)$ is.

\medskip
\noindent
$\bullet$ {\sl Perihelion shift}

Expanding $F(\Phi)$ as $F(\Phi)=\Phi+{1\over2}a_2\Phi^2+... $ 
with $\Phi=-{\cal M}/r$, a short calculation, copied from
 the standard one in GR, yields
 $\Delta\omega=-{1\over6}(1+a_2)\Delta\omega_{\rm GR}$, 
that is,  
$\beta=(1+a_2)/2$ where  $\beta$ is the PPN parameter defined
 by $(2\gamma-\beta+2)/3={\Delta\omega/\Delta\omega_{\rm GR}}$.
 Again, this prediction does not involve time measurements.
Since the least unsatisfactory of Nordtsr\"om's theories 
corresponds to $F(\Phi)=\Phi$, {\sl i.e.} $a_2=0$ (see below), 
the prediction for the perihelion shift differs from that of GR, 
$\beta=1/2$ instead of $\beta=1$, and hence is ruled out by observations.

\medskip
\noindent
$\bullet$ {\sl Geodetic precession} 

At lowest order the angular velocity of the precession of an
 accelerated spin is given by Thomas' formula which, in the 
field $\Phi=-{\cal M}/r$ where ${\cal M}\equiv{GM/c^2}$, 
reads: $\omega_{\rm Thomas}
=-{1\over2}(c/r)\left({\cal M}/R\right)^{3/2}\left({R/r}\right)^{3/2}$
 (at lowest order: $F\approx\Phi={\cal M}/r\ll1$ so that 
$\bar T\approx T)$. This is minus one third the geodetic 
precession predicted by GR and measured by GPB.

\subsection*{4. Redshifts in Nordstr\"om's gravity}

We shall here compute them using the Weyl-postulate (\ref{wtime}). For the alternative view in terms of varying masses, see \cite{Veiled}.

Consider an observer at rest at $P_0$ who receives some information from an emitter at point $P$ (this can be an explosion in a galaxy or a message from a spacecraft). Let the duration of the delivery of this information, that is, the (Weyl)-time interval between the beginning and end of reception, be $\Delta\bar\tau_{rec}$, as measured by $P_0$ (this means that his clock has ticked, 
say, $N$ times). 

The time it takes to obtain the same information at $P_0$ 
(figuratively speaking in the case of an explosion!) is 
a priori different: $\Delta\bar\tau_0$ (the clock of the 
observer at $P_0$ clicks $N_0$ times).

The red (or blue) shift $z$ is defined as 
$1+z={\Delta\bar\tau_{rec}/\Delta\bar\tau_0}$.

Now, since the clocks are the same at $P$ and $P_0$ (the
 astronaut in the spacecraft has the same wristwatch than
 his colleague on Earth),  the duration of  the message will
 be the same at $P$: $\Delta\bar\tau_0=\Delta\bar\tau$. 

To relate now $\Delta\bar\tau_{rec}$ and $\Delta\bar\tau$, we use
 the fact that time intervals are related to Minkowski time 
$T$ by (\ref{wtime}), so that: 
$\Delta\bar\tau_{rec}=(1+F_0)\Delta T_{rec}$ ($P_0$ is at rest) 
and  $\Delta\bar\tau=(1+F)\sqrt{1-{V^2/c^2}}\Delta T_{em}$ 
where $V^\alpha={dX^\alpha/dT}$ is the 3-velocity of 
the galaxy or spaceship.
 
If, finally,  information from $P$ to $P_0$ is transmitted at
 the speed of light then 
$\Delta T_{rec}=\left(1-{V/c}\right)\Delta T_{em}$
 (this the standard Doppler effect, supposing the motion
 of the emitter is along the line of sight; $V$ is negative 
if the emitter recedes from the observer).

Therefore, all in all:
\begin{equation}
1+z\equiv{\Delta\bar\tau_{rec}\over\Delta\bar\tau_0}
={\Delta\bar\tau_{rec}\over\Delta\bar\tau}
={(1+F_0)\Delta T_{rec}\over(1+F))\sqrt{1-{V^2\over c^2}}\Delta T_{em}}
={1+F_0\over1+F}\sqrt{1-{V\over c}\over1+{V\over c}}\,.
\label{genredshift}
\end{equation}
If the emitter is at rest ($V=0$) the redshift observed 
at $P_0$ tends to infinity as $1/(1+F)$ if $1+F\to0_+$.

\subsection*{5. Nordstr\"om's theories and (basic) cosmology}

Consider a gravitational potential $\Phi$ which, in some Minkowskian 
frame $X^\mu$, depends on time $T$ only. Then, as can easily be seen 
from the equation of motion (\ref{ptmotion}),
test particles (``galaxies") can be at rest. 

The light emitted by one of these galaxies at rest at $P$  travels
  along the Minkowskian cones to the observer, himself at rest
 at $P_0$. Thus, if two photons representing, say, a given 
atomic transition, are emitted within $\Delta T_{em}$ at $P$
 they will be received at $P_0$ within the same Minkowskian 
time interval $\Delta T_{rec}=\Delta T_{em}$. However, 
since, following (\ref{wtime}),  time is represented
 by $\bar\tau=(1+F)T$, and not by $T$, a redshift will be 
observed, given by (\ref{genredshift}) with $V=0$:
\begin{equation}
1+z={1+F_0\over1+F}\,.
\label{redshift}
\end{equation}

Let us now solve equations (\ref{eom}) when matter is a perfect fluid with 
stress energy-tensor given by (\ref{Tperf}).
Equation (\ref{eom}b) becomes
\begin{equation}
{\dot{\bar\epsilon}\over\bar\epsilon}
=-{3\dot F\over 1+F}(1+w)\qquad\hbox{with}\qquad
 w\equiv {\bar p\over\bar\epsilon}
\label{eneq}
\end{equation}
whose solution is, when $w=const$
\begin{equation}
\bar\epsilon\propto (1+F)^{-3(1+w)}\propto (1+z)^{3(1+w)}\,.
\label{epsilon}
\end{equation}
For $w>-1$ the energy density of matter diverges if there is
 a moment of infinite redshift in the history of the universe.

As for Equation (\ref{Phieq}) which gives the time evolution of the
 gravitational field, it becomes
\begin{equation}
\ddot\Phi={4\pi G\over c^4}{dF\over d\Phi}(1-3w)\bar\epsilon\,.
\label{ddotPhi}
\end{equation}
If we choose $F=\Phi$ as a example then Eqs.~(\ref{epsilon}-\ref{ddotPhi}) 
possess the following power-law solution
if $w>{1/3}$ or $w<-{2/3}$:
\begin{equation}
a(t)\equiv (1+\Phi)\propto t^{2\over4+3w}
\qquad\Longrightarrow\qquad\bar\epsilon\propto t^{-{6(1+w)\over4+3w}}\,.
\label{cossol}
\end{equation}
For $w>-1$,  $T=0$ is a ``Big-Bang":  matter density is infinite
 and then ever decreases; as for the ``scale factor"
$a$ it vanishes at $T=0$ and then ever increases and
the cosmological redshift (\ref{redshift}) becomes since $F=\Phi$:
\begin{equation}
1+z={1+\Phi_0\over1+\Phi_P}={a_0\over a}\,.
\end{equation}
The Euclidean 3-plane of the Big-Bang $T=0$ where the energy density 
diverges and the scale factor vanishes is thus an infinite redshift surface.

This ``Big-Bang" is interpreted, when one adheres to postulate (\ref{wtime}),
 as due to the fact that Weyl-time $\bar\tau$ stops running. In one adheres 
to the alternative postulate (\ref{wmass}), the Big-Bang  is due to the fact that
 the effective inertial masses of all particles go to zero 
(see \cite{Veiled}).\footnote{For 
$w={1\over3}$, $1+\Phi\propto T$ and the scale factor also vanishes.
 For $w=0$, $1+\Phi\propto\sqrt{1+const.^2T^2}$: the scale factor ``bounces" at $T=0$.}

\subsection*{6. Nordstr\"om's theories and ``black holes"}

Let us treat now the case of a gravitational potential $\Phi$ which depends only on the radial coordinate $r$ in some inertial frame $(T,r,\theta,\phi)$.

The static, spherically symmetric vacuum solution of the field equations (\ref{eom}) 
 is Newton's potential, whatever the function $F(\Phi)$:
\begin{equation}
\Phi=-{{\cal M}\over r} 
\qquad\hbox{with}\qquad {\cal M}={GM\over c^2}\,,
\end{equation}
where the integration constant $M$ is the gravitational mass of the central body. To be specific we shall restrict ourselves to the case when $F=\Phi$:
\begin{equation}
1+F=1-{{\cal M}\over r}\,.
\end{equation}

Let us first consider a static emitter in that field which sends light 
signals to infinity. These signals will be seen as all the more redshifted 
as the emitter is closer to $r={\cal M}$. 
More precisely, see Eq.~(\ref{genredshift})
\begin{equation}
1+z={1\over 1-{{\cal M}\over r}}\,.
\end{equation}
Thus $r={\cal M}$ is a surface of infinite redshift and can be called a ``horizon".

In order now to try to understand what happens if the emitter is beyond
 the horizon ($r<{\cal M}$) let us return to Eq.~(\ref{wtime}) and (\ref{wmass}).
 When $(1+F)$ becomes negative, (\ref{wtime}) tells us that Weyl-time, 
as measured by the emitter, runs backwards compared to the time $T$ measured 
at infinity. As for (\ref{wmass}) it tells us that its effective inertial
 mass becomes negative. Supposing that such behaviours can be
 reinterpreted \`a la Feynman-Dirac we impose that Weyl-time keeps 
running forwards or, equivalently, that masses remain positive
 (this amounts to replace $(1+F)$ by its absolute value), and (somewhat arbitratily) decree
 that all particles have turned into antiparticles.
 Thus, the emitter, if beyond the horizon, is an ``anti-emitter". 
As for the observed shift it will then be given by
\begin{equation}
1+z={1\over |1-{{\cal M}\over r}|}\,.
\end{equation}
It decreases from infinite redshift when $r\to{\cal M}_-$ 
to zero if the emitter is at the origin.

Let us consider now an observer falling radially towards $r=0$. 
Its equation of motion is given by (\ref{ptmotion}) whose first integral is
\begin{eqnarray}
\left({dr\over cd\tau}\right)^2={(1+F_{in})^2-(1+F)^2\over(1+F)^2}
\label{drdtau}
\end{eqnarray}
where $F_{in}$ is the value of $F(\Phi)$ at $r=r_{in}$ where the initial velocity is zero.
 In terms of Weyl's time it reads
\begin{equation}
\left({dr\over cd\bar\tau}\right)^2={(1+F_{in})^2-(1+F)^2\over(1+F)^4}\,.
\label{drdbar}
\end{equation}
In terms of Minkowkian time $T$ this becomes
\begin{eqnarray}
\left({dr\over cdT}\right)^2\equiv V^2={(1+F_{in})^2-(1+F)^2\over(1+F_{in})^2}\,.
\label{drdT}
\end{eqnarray}

The trajectory of the falling emitter when $F=\Phi=-{\cal M}/r$,
 is easily obtained by integration of  (\ref{drdtau}-\ref{drdT}).
 In Minkowskian time $T$ it starts at $T=0$ at $r=r_{in}>{\cal M}$ 
with zero velocity, reaches the speed of light when crossing $r={\cal M}$,
 slows down, reaches the turning point
 $r_-={\cal M}r_{in} /(2 r_{in}-{\cal M})$ in a finite time,
 and crosses back the surface $r={\cal M}$, again at the velocity of light. 

 The scenario is the same when described in terms of proper time  $\tau$ or 
Weyl's time $\bar\tau$, even if   the proper velocities ${dr/d\tau}$
 and ${dr/d\bar\tau}$ go to infinity when $r\to {\cal M}$. 
As measured by the  wrist-watch of the emitter and if one adheres
 to (\ref{wtime}) the round trip from $r_{in}$ to inside the horizon 
and back takes a finite amount of proper-time $\bar\tau$, as one finds 
by integration of (\ref{drdbar}), and nothing special happens when crossing it.
 The same holds true if one adheres to (\ref{wmass}): the round trip  takes a 
finite amount of proper-time $\tau$, 
as can be seen by integration of (\ref{drdtau}).

Let us suppose now that the emitter sends light signals to the
 observer at rest at infinity.
Formula (\ref{genredshift}) gives the redshidt  as a function
 of the position 
$r$ of the falling emitter when $V<0$ and $F=\Phi=-{{\cal M}/r}$ as:
\begin{equation}
1+z={(1-{{\cal M}\over r_{in}})
+\sqrt{(1-{{\cal M}\over r_{in}})^2
-(1-{{\cal M}\over r})^2}\over\left(1-{{\cal M}\over r}\right)^2}\,.
\label{infobs}
\end{equation}
Thus the  light signals  will be observed as infinitely 
redshifted at $P_0$ when the emitter approaches $r={\cal M}$: 
 signals sent by infalling matter die out when it reaches 
the horizon. The rate at which $z$ grows is enhanced by a
 factor $(1-{{\cal M}/r})$ compared to the static case 
because the emitter reaches the speed of light at horizon crossing.

What happens during the time the emitter is beyond the horizon? 
As we argued when considering static emitters, it should presumably be 
considered as an ``anti-emitter" by the distant observer 
and $(1-{{\cal M}/r})$ must be replaced by $|(1-{{\cal M}/r})|$. 
This does not change formula (\ref{infobs}) as long as $V$ remains negative: 
the observed redshift decreases from infinity to
 $1+z=(1-{\cal M}/r_{in})^{-1}$ 
at the turning point.\footnote{More precisely, taking into account the fact
 that measuring instruments have a finite resolution in frequencies,
 the observer at infinity looses contact with the emitter for a while.} 
After the turning point, $V$ is positive and the redshift
 formula (\ref{genredshift}) becomes
\begin{equation}
1+z={(1-{{\cal M}/r_{in}})
-\sqrt{(1-{{\cal M}/r_{in}})^2
-(1-{{\cal M}/r})^2}\over\left(1-{{\cal M}/r}\right)^2}\,.
\end{equation}
At horizon crossing, the redshift of $z$ due to the gravitational field is compensated by the blueshift due to the fact that the emitter reaches the speed of light and the combined effect results in a finite redshift at $r={\cal M}$:
\begin{equation}
1+z={1\over2\left(1-{{\cal M}/r_{in}}\right)}\,.
\end{equation}

 A caveat is however in order. Since (Weyl)-time is related to time at infinity by $d\bar\tau=(1-{\cal M}/r)dT$, it runs backwards when the emitter is inside the horizon. Alternatively its effective mass $m(1-{\cal M}/r)$ is negative. Therefore, as we have argued, the emitter suddenly becomes an ``anti-emitter" at horizon crossing.
This probably leads to quantum instabilities.

\section{``Jordan representation"  of Nordstr\"om's theories}

\subsection*{1. Conformal transformation of the field equations}

As was already known to Einstein and Fokker \cite{EinFok},
Nordstr\"om's theories of gravity can be turned into metric theories. Indeed if we introduce
\begin{equation}
\bar g_{\mu\nu}=(1+F)^2\,\ell_{\mu\nu}\,,
\end{equation}
then the equations of motion (\ref{eom}) can be recast as
\begin{equation}
\left\{\begin{aligned}
&\bar R={24\pi G\over c^4}
\left({dF\over d\Phi}\right)^2\bar T_{\rm m}
-{6\over1+F}{d^2F\over d\Phi^2}\bar g^{\mu\nu}\partial_\mu\Phi\partial_\nu\Phi\,,
\cr
&\bar D_j\bar T^{\mu\nu}_{\rm m}=0\,,
\end{aligned}\right.
\end{equation}
where $\bar R$ is the scalar curvature of the conformally flat metric $\bar g_{\mu\nu}$
 and where $\bar T_{\mu\nu}$ is the stress-energy tensor of matter minimally
 coupled to the metric $\bar g_{\mu\nu}$.
 Thus the action and stress-energy tensor for particles are
\begin{equation}
S_{\rm m}=-\sum mc^2\int d\bar\tau\qquad,\qquad \bar T^{\mu\nu}_{\rm m}
=\sum mc\int {\bar u^\mu \bar u^\nu\over\sqrt{-\bar g}}
\delta_4[x^\lambda-x^\lambda(\tau)]d\bar\tau\,,
\end{equation}
with $\bar u^\mu={dx^\mu\over d\bar\tau}$ and
 $\bar g_{\mu\nu}\bar u^\mu\bar u^\nu=-c^2$.
 The action and stress-energy tensor for a scalar field $\Psi$ are
\begin{equation}
\left\{\begin{aligned}
S_{\rm m}&=-\int {d^4x\over c}\sqrt{-\bar g}
\left[{1\over2}\bar g^{\mu\nu}\partial_\mu\Psi\partial_\nu\Psi+V(\Psi)\right]\,,
\cr
\bar T_{\mu\nu}^{\rm m}&=\partial_\mu\Psi\partial_\nu\Psi
-\bar g_{\mu\nu}\left({1\over2}\partial_\rho\Psi\bar\partial^\rho\Psi+V(\Psi)\right)\,.
\end{aligned}\right.
\end{equation}

Finally the stess-energy tensor for a perfect fluid is~: 
\begin{equation}
\bar T_{\mu\nu}^{\rm m}
=(\bar\epsilon+\bar p){\bar u_\mu\bar u_\nu\over c^2}+\bar p\,\bar g_{\mu\nu}\,.
\end{equation}

In this ``Jordan representation", $T$ (called ``Minkowskian time" in the previous section) becomes a mere coordinate time with no special significance and $\bar\tau$ (previously called ``Weyl's time") measures the length of worldlines by means of the metric $\bar g_{\mu\nu}$ and is postulated to represent time as measured along the worldline. Therefore all predictions are the same, whether one uses the previous, ``Einstein" representation or this ``Jordan" one, although the descriptions may vary, see below some examples.

In the ``Jordan representation"  the special status of the choice 
\begin{equation}
F(\Phi)=\Phi
\end{equation}
 is also manifest. In that case indeed the equations of the theory reduce to
 (see \cite{DamourEsp})
\begin{equation}
\bar R={24\pi G\over c^4}\bar T_{\rm m}\qquad,\qquad 
\bar D_\nu\bar T^{\mu\nu}_{\rm m}=0\qquad,\qquad\bar C_{\mu\nu\rho\sigma}=0\,,
\label{barReq}
\end{equation}
where the vanishing of the Weyl tensor $\bar C_{\mu\nu\rho\sigma}$ imposes the 
metric to be conformally flat. Equations (\ref{barReq}) share with 
Einstein's equations the fact that they are purely geometrical and second order. 
Hence the claim that  Nordstr\"om's theory with $F=\Phi$ embodies 
the strong equivalence principle \cite{DamourEsp}.

\subsection*{2. Cosmology and ``black holes" revisited}

When working in the Jordan representation, the cosmological solution (\ref{cossol})
is interpreted as an ``expansion of the universe", test particles (galaxies) are 
said to be at rest in the ``comoving" frame $x^\mu\equiv (t,r,\theta,\phi)$
 and the redshift is interpreted as due to the expansion of the universe. 
As for the Big-Bang it is not only a surface where the energy density of
 matter diverges, it is also a curvature singularity.

Similarly the infinite redshift surface at $r={\cal M}$ in a static spherically symmetric gravitational field discussed in II.6   is also a curvature singularity of the Jordan metric,
\begin{equation}
d\bar s^2=\left(1-{{\cal M}\over r}\right)^2(-dT^2+dr^2+r^2d\Omega^2)\,.
\end{equation}
The square of the Ricci tensor for example is given by
\begin{equation}
\bar R_{\mu\nu}\bar R^{\mu\nu}
={4{\cal M}^2({\cal M}^2-4{\cal M}r+6r^2)\over(r-{\cal M})^8}\,.
\end{equation}
Now the motion of a test particle is the geodesic equation, 
${\bar D\bar u^\mu/d\bar\tau}=0$ with
 $\bar g_{\mu\nu}\bar u^\mu\bar u^\nu=-c^2$, 
which is identical to (\ref{ptmotion}).
Therefore, as discussed in II.6 a freely falling object, can cross 
the curvature singularity at $r={\cal M}$. Note also that $r=0$ is not
a curvature singularity: all components of $R^\mu_\nu$
 are well-behaved there.

One can therefore argue that the Jordan spacetime describing the gravitational
 field of a  static spherically symmetric point-like object is geodesically 
incomplete, its completion being the Minkowski spacetime of the Einstein
 representation.

However, as we discussed it in II.6, if particles beyond $r={\cal M}$ are 
considered as antiparticles then a quantum description of $r={\cal M}$ 
 becomes necessary.

\section{Conclusions}

As we have briefly recalled, Nordstr\"om's theory of gravity is 
observationally ruled out, but is interesting as a playground where to 
probe various aspects of metric theories of gravity. 
By presenting some gedanken experiments in the field of a compact object 
we have seen in particular that a  spurious curvature singularity
 (in that geodesics can cross it) can be removed and the manifold be
 extended by means of a conformal transformation, at least at the classical level.

\begin{acknowledgments}

N.D. thanks Gilles Esposito-Far\`ese for discussions and
the Yukawa Institute for its hospitality when this work was completed. 
She also acknowledges financial support from the CNRS-JSPS contract 24600.
M.S. is supported by 
Korea Institute for Advanced Study under the KIAS Scholar program,
by the Grant-in-Aid for the Global COE Program at Kyoto University,
``The Next Generation of Physics, Spun from Universality and Emergence''
from the Ministry of Education, Culture, Sports, 
Science and Technology (MEXT) of Japan,
by JSPS Grant-in-Aid for Scientific Research (A) No.~21244033,
and by JSPS Grant-in-Aid for Creative Scientific Research No.~19GS0219.
This work was done as part of research activities at
the long-term workshop ``Gravity and Cosmology 2010 (GC2010)''
 (YITP-T-10-01) and the YKIS symposium ``Cosmology -- The Next Generation --'' (YKIS2010).
\end{acknowledgments}

\appendix

\section{The coupling of gravity to matter in Nordstr\"om's theories}

If matter  is a scalar field $\Psi(x^\mu)$ it action can {\sl a priori} be taken
to be the sum of a kinetic and a potential term coupled to gravity as
\begin{equation}
S_{\rm m}
=-\int {d^4x\over c}\sqrt{-\ell}
\left[G_1(\Phi){1\over2}\ell^{\mu\nu}\partial_\mu\Psi\partial_\nu\Psi
+G_2(\Phi)V(\Psi)\right]\,,
\end{equation}
where $G_1$ and $G_2$ have to be chosen appropriately. 
The equation of motion for $\Phi$ is
\begin{equation}
\square\Phi={4\pi G\over c^4}
\left({1\over2}{dG_1\over d\Phi}\partial_\rho\Psi\partial^\rho\Psi
+{dG_2\over d\Phi}V(\Psi)\right)\,.
\end{equation}
The stress-energy tensor of the field $\Psi$ is, on the other hand~:
\begin{eqnarray}
T_{\mu\nu}^{\rm m}=G_1\partial_\mu\Psi\partial_\nu\Psi
-\ell_{\mu\nu}\left({1\over2}G_1\partial_\rho\Psi\partial^\rho\Psi+G_2V(\Psi)\right)\,.
\end{eqnarray}
If one now imposes that the source for gravity be related to the stress-energy
 tensor of matter (a requirement made by Einstein \cite{Einstein}, 
quoted in \cite{Ravndal}), 
so that the field equations have the universal form (\ref{Phieq}), we see that the
 two functions $G_1$ and $G_2$ are  related to $F$ thus~: 
$G_1=(1+F)^2$~; $G_2=(1+F)^4$ so that $S_{\rm m}$ and $T^{\rm m}_{\mu\nu}$ become
\begin{eqnarray}
S_{\rm m}
&=&-\int {d^4x\over c}(1+F(\Phi))^4\sqrt{-\ell}
\left[{\ell^{\mu\nu}\over (1+F(\Phi))^2}
{1\over2}\partial_\mu\Psi\partial_\nu\Psi+V(\Psi)\right]\,,
\\
T_{\mu\nu}^{\rm m}&=&(1+F(\Phi))^2
\left(\partial_\mu\Psi\partial_\nu\Psi
-{1\over2}\ell_{\mu\nu}\partial_\rho\Psi\partial^\rho\Psi
-(1+F(\Phi))^2\ell_{\mu\nu}V(\Psi)\right)\,.
\end{eqnarray}

When matter now is a perfect fluid for which an action is awkward to define,
 one imposes the equations of motion to be (\ref{Phieq}) and (\ref{ptmotion}) where, now, 
$T_{\mu\nu}^{\rm m}$ is the stress-energy tensor of the fluid interacting
 with the gravity field $\Phi$. In analogy with the procedure followed 
when treating a scalar field one may start from the following ansatz~:
\begin{eqnarray}
T_{\mu\nu}^{\rm m}=\left[\bar \epsilon H_1(\Phi)
+\bar  pH_2(\Phi)\right]{u_\mu u_\nu\over c^2}+\bar pH_2(\Phi)\ell_{\mu\nu}\,,
\end{eqnarray}
where the functions $H_1$ and $H_2$ must be chosen appropriately and 
where $\bar \epsilon$ and $\bar p$ are the energy density and pressure of the fluid.
A way to restrict the functions $H_1$ and $H_2$ is first to impose that a
 radiation fluid ($\bar\epsilon=3\bar p$) does not source gravity and is not 
influenced by it (this is in keeping with the fact that light follows light
 cones in Minkowski spacetime even in the presence of gravity).
 This requirement yields $H_2=H_1$. At this stage and using, admittedly, 
the insight shown in  \cite{EinFok},
we see that if one chooses $H_1(\Phi)=(1+F(\Phi))^4$ and if one introduces 
the auxiliary metric $\bar g_{\mu\nu}=(1+F(\Phi))^2\ell_{\mu\nu}$
 (which is nothing but the Jordan metric), then the equations of 
motion (\ref{ptmotion}) reduce to equations of motion for a free fluid,
\begin{eqnarray}
\bar D_\nu\bar T^{\mu\nu}_{\rm m}=0
\quad\Longleftrightarrow\quad
\left\{\begin{aligned}
{d\bar\epsilon\over d\bar\tau}&=-(\bar\epsilon+\bar p)
\bar D_\nu\bar u^\nu\cr
{\bar D\bar u_\mu\over d\bar\tau}
&=-{c^ 2\over\bar\epsilon+\bar p}\left(\partial_\mu\bar p
+{\bar u_\mu\bar u^\nu\over c^2}\partial_\nu\bar p\right)\,,
\end{aligned}\right. 
\end{eqnarray}
where $\bar D_\mu$ is the covariant derivative associated with 
$\bar g_{\mu\nu}$,  $\bar u^\mu={dx^\mu/d\tau}$, 
$\bar g_{\mu\nu}\bar u^\mu\bar u^\nu=-c^2$ and 
$T_{\mu\nu}^{\rm m}
=(\bar\epsilon+\bar p){\bar u_\mu\bar u_\nu/c^2}
+\bar p\bar g_{\mu\nu}$.
 Hence, with the above choice for $H_1$,  there exists a ``freely falling frame" where gravity is
 effaced and where, in accordance with the Einstein equivalence principle, local physics apply.

 \vfill\eject

\end{document}